\documentclass[preprint,12pt]{elsarticle}

\usepackage{amssymb}
\usepackage{amsmath}
\usepackage{svg}
\usepackage{caption}

\begin{document}

\begin{frontmatter}



\title{Understanding the Mechanisms Behind Structural Influences on Link Prediction: A Case Study on FB15k-237}


\author[1]{ Xiaobo Jiang} 
\author[1]{ Yadong Deng} 
\cortext[cor1]{\textit{Yadong Deng}. 
               E-mail: 202321011978@mail.scut.edu.cn}

\affiliation[1]{organization={South China University of Technology},
            city={Guangzhou},
            postcode={510641}, 
            country={China}}

\begin{abstract}
FB15k-237 mitigates the data leakage issue by excluding inverse and symmetric relationship triples, however, this has led to substantial performance degradation and slow improvement progress.  Traditional approaches demonstrate limited effectiveness on FB15k-237, primarily because the underlying mechanism by which structural features of the dataset influence model performance remains unexplored. To bridge this gap, we systematically investigate the impact mechanism of dataset structural features on link prediction performance. Firstly, we design a structured subgraph sampling strategy that ensures connectivity while constructing subgraphs with distinct structural features. Then, through correlation and sensitivity analyses conducted across several mainstream models, we observe that the distribution of relationship categories within subgraphs significantly affects performance, followed by the size of strongly connected components. Further exploration using the LIME model clarifies the intrinsic mechanism by which relationship categories influence link prediction performance, revealing that relationship categories primarily modulate the relative importance between entity embeddings and relationship embeddings and relationship embeddings, thereby affecting link prediction outcomes. These findings provide theoretical insights for addressing performance bottlenecks on FB15k-237, while the proposed analytical framework also offers methodological guidance for future studies dealing with structurally constrained datasets.
\end{abstract}




\end{frontmatter}



\section{Introduction}
\label{sec1}

The evolution of modern artificial intelligence systems critically depends on high-quality datasets. Datasets not only determine models' learning capabilities and predictive accuracy but also drive innovations in interdisciplinary research paradigms \cite{ref1,ref2,ref3,ref4,ref5,ref6}. A prominent trend in dataset design involves eliminating redundant triples, such as inverse and symmetric relationships, to mitigate data leakage and facilitate genuine reasoning rather than simple memorization \cite{ref7}. 
FB15k-237 is a typical example of this approach. Derived from the FB15K dataset \cite{ref8}, FB15k-237 reduces data leakage by removing redundant relationship triples. However, its stringent construction criteria have led to notably lower performance across various models and slow performance improvements, as illustrated in Figure~\ref{fig_1}. Under the same model settings and training scale, the highest Mean Reciprocal Rank (MRR) achieved on FB15k-237 is only 0.42, significantly lower than those achieved on similar datasets such as FB15k (0.86) \cite{ref9} and WN18RR (0.74) \cite{ref10}. Consequently, FB15k-237 has increasingly been recognized as a more challenging benchmark for evaluation \cite{ref11}.

\begin{figure}[!t]
\captionsetup{aboveskip=0pt,  
            belowskip=-16pt}  
\centering
\includegraphics[width=\linewidth]{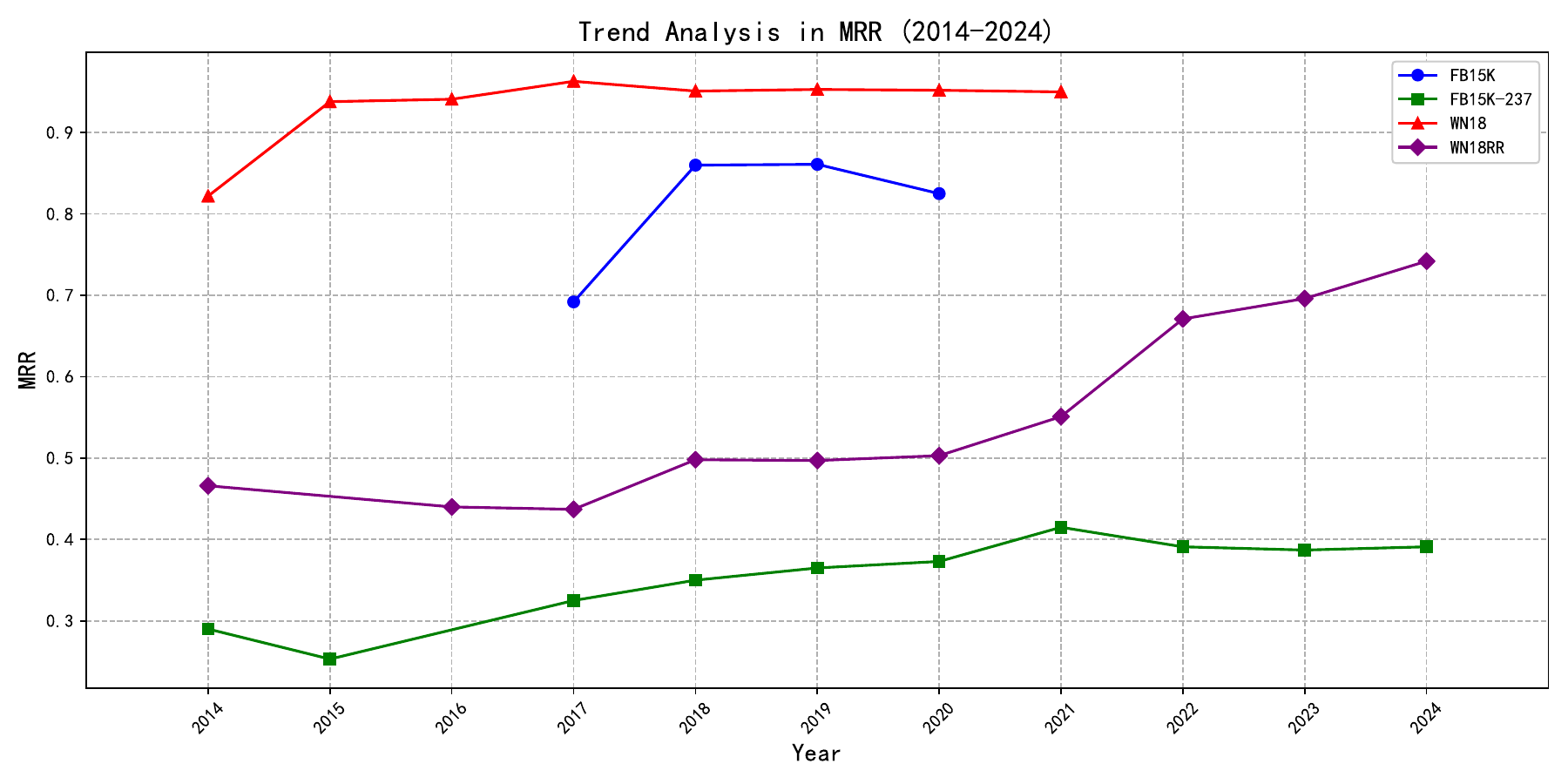}
\caption{Curve of MRR performance changes across different datasets}
\label{fig1}
\end{figure}

Although traditional research typically attributes performance gains to optimizing complex entity relationships, semantic associations, and model parameters, these methods show limited effectiveness on FB15k-237. For example, early translation-based models like TransE have been extended by more advanced variants such as TransD \cite{ref12} and TransR \cite{ref13} to better capture complex relationship patterns. Similarly, bilinear models including DistMult \cite{ref14}, ComplEx \cite{ref15}, and TuckER \cite{ref16} utilize matrix and tensor decomposition to model semantic associations. Neural-network-based approaches, such as ConvE \cite{ref17}, which leverages convolutional architectures, and HittER \cite{ref18}, employing Transformer architectures, have also been extensively explored. Furthermore, graph neural network models like NBFNet \cite{ref19}, as well as techniques involving hyperparameter tuning and diversified loss functions, have been investigated \cite{ref20,ref21,ref22}. However, methods successful on conventional benchmarks fail to deliver expected performance improvements on FB15k-237. Performance curves across different datasets (shown in Figure~\ref{fig1}) suggest that the traditional factors affecting model effectiveness do not entirely apply within its more restrictive structure.

Since the construction of FB15k-237 only modified the dataset's structure without changing the entity and relation information, the slow performance improvement may primarily relate to its structural features. We hypothesize that the fundamental performance bottleneck arises from an unexplored mechanism concerning how structural features of FB15k-237 impact model performance. Therefore, this study systematically investigates the mechanism through which dataset structural features influence link prediction performance on FB15k-237. To this end, we propose a randomized subgraph sampling method, utilizing a heuristic breadth-first search (BFS) strategy, to extract connected subgraphs of varying sizes and structural features from the large-scale multi-relational graph. Through correlation and sensitivity analyses conducted on these subgraphs, we quantify the influence of different structural features on model performance. Our results indicate that the distribution of relationship categories within subgraphs has the most significant impact, followed by the size of strongly connected components. Consistent observations across multiple models further verify the universality of these findings, confirming that these influences originate intrinsically from dataset properties rather than being model-specific. Finally, by employing the LIME model, we reveal the pathways through which relationship categories affect model performance. Specifically, we demonstrate that relationship categories primarily influence link prediction by modulating the relative importance of entity embeddings and relationship embeddings.

In summary, this paper reveals the key structural features of the dataset that affect performance, quantifies their impact, and clarifies the pathways through which these features affect link prediction outcomes. Consistent results across multiple models further validate our findings. These insights provide a theoretical basis for improving performance on the FB15k-237 dataset, offering crucial support for future research aiming to enhance the learning capability of complex relationship categories, optimize resource allocation strategies, and improve embedding representations. Additionally, the analytical approach proposed in this study can be generalized to other datasets with performance bottlenecks, establishing a universal analytical framework for investigating and optimizing structural features. Consequently, this work contributes positively to the ongoing advancement and broader application of knowledge graph link prediction technologies.

The structure of the paper is as follows: Chapter 1 introduces the research background and objectives of the study; Chapter 2 presents the relevant Knowledge Graph Embedding (KGE) models and previous studies on mechanism features; Chapter 3 discusses the experimental methods, including the random sampling method, the Gini index to measure data distribution imbalance, and the principles of using LIME for explanatory mechanism analysis; Chapter 4 is the experimental section, where the proposed ideas are validated; Chapter 5 summarizes the contributions of this work.

\section{Related work}
\label{sec2}
\subsection{KGE model}
\label{subsec1}
Knowledge graph completion (KGC) aims to identify missing interaction relationships between entities, thereby addressing the problem of incomplete knowledge graphs. Knowledge graph embedding (KGE) models infer new facts by mapping the elements of a knowledge graph into high-dimensional vectors and defining a scoring function. Typically, KGE models follow four steps \cite{ref24}: (1) Random Initialization: Entities and relation vectors are randomly initialized; (2) Scoring Function: A scoring function is defined to measure the plausibility of a triple; (3) Interaction Mechanism: An interaction mechanism is designed to model the interactions between entities and relations and calculate the matching score of the triple; (4) Training Strategy: Strategies such as negative sampling and regularization are employed to maximize the confidence of triples and train the KGE model.

Based on this process, various models have been proposed. The earliest of these is the translation-based TransE model, which is simple and efficient, but has limited capability in representing complex relationships. This limitation motivated subsequent research to further extend the modeling capacity of knowledge graph embeddings. For example, to address the shortcomings of TransE in handling complex relationship categories, researchers proposed TransH \cite{ref25} and TransR \cite{ref14}. TransH introduces relation-specific hyperplanes, enabling the same entity to have different projected representations under different relations, thus better handling complex relationships. TransR further proposes embedding entities and relations into separate vector spaces, where projection matrices are used to project entities into relation-specific spaces, solving the problem of insufficient representation when entities and relations share the same space. Subsequently, RotatE \cite{ref16} defines each relation as a rotation operation in the complex vector space, representing the relation as a rotation from the head entity to the tail entity. This design allows RotatE to effectively model symmetric, antisymmetric, and many-to-many relationships, addressing the limitations of TransE in handling such relationship categories.

In addition to the aforementioned translation-based models, there are also semantic models that use trilinear products to measure the semantic similarity between entities and relations. For example, DistMult uses a bilinear model to capture the interaction between entities and relations, with its scoring function based on a trilinear product. This design allows DistMult to perform well in modeling symmetric relations but struggles to accurately represent antisymmetric relations. ComplEx extends the embedding representations of entities and relations from real space to complex space, enabling better representation of more complex relationship categories such as cyclic relations and many-to-many relationships. Additionally, some models leverage deep learning techniques to capture more complex patterns in knowledge graphs. For instance, ConvE \cite{ref18}, based on convolutional neural networks, captures local feature interactions between entity and relation embeddings, effectively learning diverse combinations of entities and relations. Similarly, CapsE \cite{ref26} introduces capsule networks into the knowledge graph embedding framework to capture more fine-grained feature interactions between entities and relations, demonstrating superior performance in modeling complex relationships.

Despite the numerous efforts by researchers to improve the performance of KGE models in link prediction tasks, current experimental results seem to have encountered a bottleneck, especially in complex datasets such as the one addressed in this study. This suggests that improving performance through optimization of the model's structure and parameters alone has reached a limited effect. Therefore, it is urgently needed to analyze the inherent mechanisms of models from the perspective of dataset structural features and explore the potential causes of performance bottlenecks, providing direction and insights for subsequent model improvements.

\subsection{Mechanism study}
\label{subsec2}
In recent years, significant progress has been made in the research of Knowledge Graph Embedding (KGE) models; however, there remains a gap in understanding the key factors that influence model performance. Some existing studies have aimed to improve embedding quality by optimizing model design, adjusting hyperparameters, and addressing data distribution issues. For instance, Oliver et al. \cite{ref27} used Sobol sensitivity analysis to assess the impact of different hyperparameters on the variance of embedding quality, aiming to identify which hyperparameters can be excluded from the search space without significantly affecting embedding quality. Zhang et al. \cite{ref12} proposed the Weighted Knowledge Graph Embedding (WeightE) method to examine the impact of long-tail distributions on model performance. Akrami et al. \cite{ref28} revealed that data leakage issues may lead to an overestimation of the performance of embedding models. In terms of KGE model interpretability, Zhang et al. \cite{ref29} introduced the Path-based Heterogeneous Link Prediction GNN explanation method (PaGE-Link) to address the lack of interpretability in GNN models for link prediction tasks. PaGE-Link can generate interpretable paths and offers scalable model capabilities. Ma et al. \cite{ref30} proposed KGExplainer to address the limitations of existing KGE model explanation methods when reasoning information is insufficient.

In the FB15k-237 dataset, after removing redundant relationships, its structural features significantly affect the learning process and performance of embedding models. However, this aspect has not been thoroughly explored. Therefore, this study aims to investigate the mechanisms by which its structural features influence embedding performance.

\section{Method}
\label{sec3}
\subsection{Subgraph Sampling Methods}
\label{subsec1}
In this paper, we propose a connected subgraph sampling method to construct subgraphs with different structural features. This method is based on a heuristic strategy that prioritizes nodes with high connectivity \cite{ref31}. It uses breadth-first search (BFS) \cite{ref32} to extract subgraphs of a specified size from large-scale multi-relational graphs. The specific method is as follows:

Given a knowledge graph \( G = (V, E) \), where the number of nodes is \( |V| \), the goal of the sampling process is to obtain a subgraph \( G_s = (V_s, E_s) \), where the number of nodes in the subgraph \( V_s \) satisfies:

\begin{equation}
|V_s| = \lceil r \cdot |V| \rceil
\end{equation}

In this method, \( r \) represents the sampling ratio, which determines the number of nodes to sample. This ratio is randomly selected from a uniform distribution within the range \([r_{\min}, r_{\max}]\).

To improve the connectivity of the sampled subgraph, the method prioritizes sampling from high-degree nodes. Specifically, the degree of a node \( v \in V \) is defined as:

\begin{equation}
\deg(v) = \left|\{ u \in V : (u, v) \in E \text{ or } (v, u) \in E \}\right|
\end{equation}

In other words, the degree of node \( v \) is the number of its direct neighbors. After sorting the nodes in descending order by degree, the top \( k \) nodes are selected to form a candidate set:

\begin{equation}
S_k = \{ v_1, v_2, \ldots, v_k \}
\end{equation}
where

\begin{equation}
\deg(v_1) \ge \deg(v_2) \ge \cdots \ge \deg(v_k)
\end{equation}

From this candidate set, a node \( v_0 \) is randomly selected as the starting point for sampling. The sampled node set \( V_s \) and the expansion queue \( Q \) are initialized as follows:

\begin{equation}
V_s = \{ v_0 \}, \quad Q = [v_0]
\end{equation}

Then, the algorithm iteratively selects the current node \( u \) from the queue \( Q \) and explores all its neighboring nodes:

\begin{equation}
N(u) = \{ v \in V : (u, v) \in E \text{ or } (v, u) \in E \}
\end{equation}

The unvisited neighbor nodes are added to the sampled set \( V_s \) and the queue \( Q \), as follows:

\begin{equation}
V_s = V_s \cup \left(N(u) \setminus V_s\right), \quad Q = Q \cup \left(N(u) \setminus V_s\right)
\end{equation}

This process repeats until the condition is satisfied:

\begin{equation}
|V_s| \ge \lceil r \cdot |V| \rceil
\end{equation}

This sampling method ensures that the sampled subgraph \( G_s \) is connected, retaining the structural features of the knowledge graph more completely, which helps in studying the local structural features of the knowledge graph.

\subsection{Gini Index}
\label{subsec2}
In this study, we use the Gini index \cite{ref33} to quantify the imbalance in the distribution of relationship categories, relation types, and node degrees. The Gini coefficient is a widely used measure for assessing the degree of imbalance in a distribution. Originally developed to measure resource allocation and economic inequality, it is also applicable to describe imbalances in other domains, such as data distributions in knowledge graphs. Mathematically, the Gini coefficient is defined as the ratio of the average absolute difference between any two data points and the mean of the data, and is expressed as:

\begin{equation}
G = \frac{\frac{1}{n^2} \sum_{i=1}^{n} \sum_{j=1}^{n} \left| x_i - x_j \right|}{2 \bar{x}},
\end{equation}

where \( n \) is the total number of samples, \( x_i \) is the \( i \)-th data point, and \( \bar{x} = \frac{1}{n} \sum_{i=1}^{n} x_i \) is the mean of the samples. A larger Gini index \( G \) indicates a more uneven distribution.

\subsection{LIME Model}
\label{subsec3}
LIME (Local Interpretable Model-agnostic Explanations) \cite{ref34} is a general-purpose model explanation method. It aims to approximate the behavior of complex models in a local region using a simple linear model, thereby revealing the decision-making process of black-box models. The basic principle is that, in specific regions of the input space, a complex model can be locally approximated by a simple linear model. The main goal of LIME is to assess the importance of input features and quantify their contribution to the model's output. The method is based on the following principle:

Assume the black-box model is

\begin{equation}
f: \mathbb{R}^d \to \mathbb{R},
\end{equation}

where the input is \( x \in \mathbb{R}^d \) and the output is the predicted value \( f(x) \). LIME generates an explanation through the following process: First, a set of perturbed samples \( \{ x_i' \}_{i=1}^{n} \) is generated in the neighborhood of the input sample \( x \). These perturbed samples are then predicted by the black-box model \( f \), obtaining the corresponding predicted values \( \{ f(x_i') \}_{i=1}^{n} \). Next, a weight function \( \pi_x(x') \) is defined, typically using a Gaussian kernel function:

\begin{equation}
\pi_x(x') = \exp\left(-\frac{\|x - x'\|^2}{\sigma^2}\right),
\end{equation}

which represents the weight of the perturbed sample \( x' \) relative to the original input \( x \), where \( \sigma \) is the bandwidth parameter.

Finally, in the space of perturbed samples, a simple linear model \( g \) is fitted using weighted least squares to minimize the following weighted error function:

\begin{equation}
g(z') = \beta_0 + \sum_{j=1}^{d} \beta_j z_j',
\end{equation}

\begin{equation}
L(f,g,\pi_x) = \sum_{i=1}^{n} \pi_x(x_i') \cdot \left(f(x_i') - g(z_i')\right)^2 + \Omega(g).
\end{equation}

Here, \( L \) is the weighted squared error, used to measure how well the linear model \( g \) fits the black-box model \( f \); \( \Omega(g) \) represents a regularization term to control the complexity of the linear explanation model \( g \) (e.g., by applying L1 regularization to limit the dimensionality of the features).

The parameters of the fitted linear model \( \{\beta_j\}_{j=1}^{d} \) represent the feature importance weights, reflecting the contribution of each feature to the model prediction \( f(x) \) near the sample \( x \), and can be used to explain the influence of the features on the model.

In this paper, LIME is used to explain the impact of different embedding dimensions on the model score. For an input triple \( (h, r, t) \), its embedding features \( x = [h, r, t] \) define the scoring function of the model \( f \), which is the prediction score of a KGE model (e.g., TransE):

\begin{equation}
f(x) = -\|h + r - t\|_1.
\end{equation}

By optimizing the objective function \( L(f, g, \pi_x) \), LIME derives the linear model parameters \( \{\beta_j\}_{j=1}^{d} \) for the embedding features \( x = [h, r, t] \), and the feature importance is computed as:

\begin{equation}
I_j = |\beta_j|,
\end{equation}

where \( I_j \) represents the importance of feature \( j \).

By classifying the importance based on embedding type (head entity, relation, tail entity), the total contribution of each embedding type to the score can be computed:

\begin{equation}
I_{\text{head}} = \sum_{j \in \text{head}} |\beta_j|, \quad I_{\text{relation}} = \sum_{j \in \text{relation}} |\beta_j|, \quad I_{\text{tail}} = \sum_{j \in \text{tail}} |\beta_j|.
\end{equation}

Thus, through the LIME model, the feature importance of head entities, tail entities, and relation embeddings can be obtained.

\section{Experiments}
\label{sec4}
In this section, this paper further investigates the impact of dataset structural features on the performance of link prediction models by analyzing several structural features of the FB15K237 dataset.

\subsection{Experimental Setup}
\label{subsec1}
In the link prediction task, different models show varying performance on several commonly used benchmark datasets. Research indicates that, regardless of whether traditional models or more advanced models introduced in recent years, the performance on the FB15k-237 dataset is generally low. Furthermore, methods that have been successful on other datasets show limited improvement on this dataset. This suggests that the structural features of the FB15k-237 dataset itself may have a significant impact on model performance. Therefore, this paper selects the FB15k-237 dataset as the primary subject of the experiments to analyze the specific impact factors of its structural features on KGE model link prediction performance. The chosen models for the experiments are representative ones: TransE, ComplEx, and RotatE, and the experiments are conducted using the open-source library OpenKE~\cite{ref35}.

The evaluation metric used in this paper is the Mean Reciprocal Rank (MRR), which represents the average inverse rank of correct entities and is used to measure the overall ranking performance of the model.

\begin{figure*}[!t]
  \centering
  \captionsetup{aboveskip=-2pt,  
                belowskip=-12pt}  
  \includegraphics[width=\linewidth]{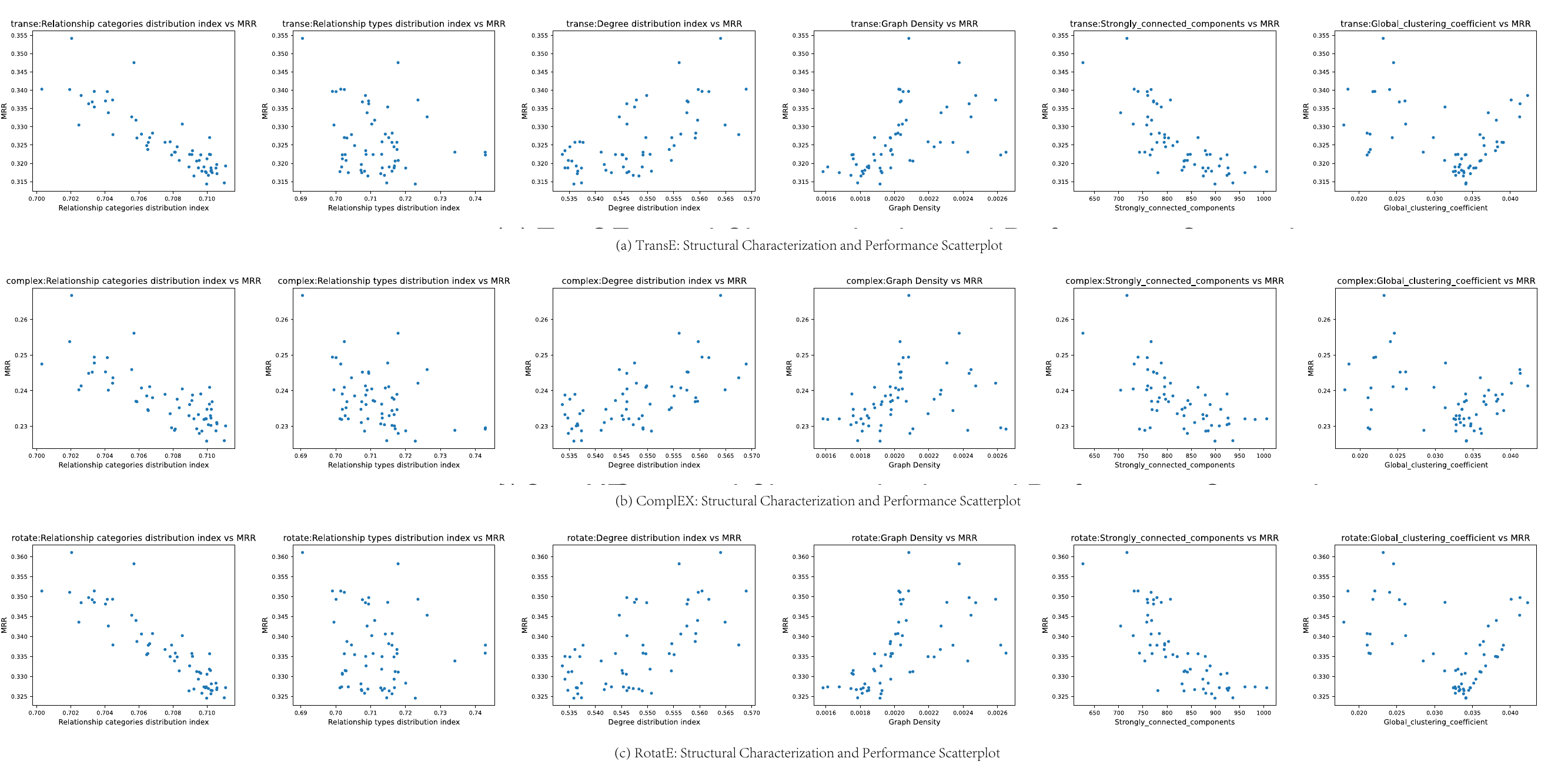}
  \caption{Scatterplot of the relationship between different structural features and MRR performance.}
  \label{fig2}
\end{figure*}

\subsection{Correlation Analysis}
\label{subsec2}
To investigate the relationship between the structural features of knowledge graphs and model performance, this paper proposes a subgraph sampling method based on a high-degree node heuristic strategy and breadth-first search (BFS). This method samples the FB15k-237 dataset to generate 60 connected subgraphs with different structural features. These subgraphs differ from the original graph in their structural features. After analyzing the dataset, this paper studies the following structural features to examine their impact on model performance:

\begin{itemize}
    \item \textbf{relationship category distribution index:} Measures the balance of relationship category distribution, calculated using the Gini coefficient. 
    \item \textbf{Relation type distribution index:}  Evaluates the imbalance in the number of triples corresponding to different relationship types, calculated using the Gini coefficient.
    \item \textbf{Degree distribution index:} Assesses the imbalance in node connectivity distribution, calculated using the Gini coefficient.
    \item \textbf{Graph density:} Represents the ratio of the actual number of edges to the maximum possible number of edges in the graph.
    \item \textbf{Number of strongly connected components:} Counts the number of strongly connected subgraphs in the graph.
    \item \textbf{Global clustering coefficient:} Measures the proportion of triangles (i.e., triples formed between nodes) in the graph. 
\end{itemize}

To study the correlation between different structural features and model performance, this paper re-trains and tests the model on each subgraph and records the model's performance (MRR) on each subgraph. After obtaining the results, scatter plots showing the relationship between subgraphs with different structural features and performance are plotted, as shown in Figure~\ref{fig2}.

Through the scatter plots, we observe the following key points: (1) The relationship category distribution index shows a clear correlation with performance, exhibiting a strong negative correlation. Specifically, the more imbalanced the relationship category distribution, the worse the model performance. An imbalanced relationship category distribution leads to a severe shortage of training samples for certain relationship categories, making it difficult for the model to effectively learn the semantic patterns of these relationships, which in turn impacts the model’s generalization ability. (2) Next, the number of strongly connected components shows a negative correlation with performance. An increase in the number of strongly connected components suggests a reduction in the overall connectivity of the graph, which may hinder the model’s ability to integrate semantic information across subgraphs, affecting inference performance and ultimately reducing the model's generalization capability. (3) The relationship type distribution index, degree distribution index, graph density, and global clustering coefficient show a certain degree of positive correlation with performance, although not as clearly. This indicates that these factors may not be the key drivers of model performance.

\begin{table*}[!t] 
\caption{Correlation Analysis Results\label{TABLE1}}
\centering
\footnotesize
\tabcolsep 6pt 
\begin{tabular*}{\textwidth}{@{\extracolsep{\fill}}lcccc}  
\hline
Structural Characteristic                    & Method   & TransE  & ComplEX & RotatE \\ 
\hline
Relationship categories distribution index   & Pearson  & -0.8795 & -0.7829 & -0.9056 \\
                                             & Spearman & -0.8777 & -0.7575 & -0.8963 \\
\hline
Relationship types distribution index        & Pearson  & -0.2514 & -0.4274 & -0.1580 \\
                                             & Spearman & -0.1671 & -0.3774 & -0.1238 \\
\hline
Degree distribution index                    & Pearson  & 0.6386  & 0.6292  & 0.6400  \\
                                             & Spearman & 0.6157  & 0.5975  & 0.6233  \\
\hline
Graph density                                & Pearson  & 0.5118  & 0.2959  & 0.5889  \\
                                             & Spearman & 0.6921  & 0.4271  & 0.7267  \\
\hline
Strongly connected components                & Pearson  & -0.7601 & -0.6364 & -0.7920 \\
                                             & Spearman & -0.7980 & -0.6143 & -0.8098 \\
\hline
Global clustering coefficient                & Pearson  & -0.3376 & -0.3283 & -0.3599 \\
                                             & Spearman & -0.1613 & -0.1417 & -0.1941 \\
\hline
\end{tabular*}
\end{table*}

To further validate these relationships, this paper employs Pearson's correlation\cite{ref36} coefficient and Spearman's rank correlation coefficient\cite{ref37} to quantify the relationship between structural features and model performance. Both Pearson's and Spearman's correlation coefficients are commonly used to detect linear relationships between variables within datasets. The results are shown in Table~\ref{TABLE1}. It can be observed that the relationship category distribution index has the most significant correlation with performance. This finding underscores the importance of the relationship category distribution in the structure of the FB15k-237 dataset, indicating that a more balanced relationship category distribution corresponds to better subgraph performance. It is important to emphasize that by conducting correlation experiments on different models, we obtained the same result: the relationship category distribution within the subgraph showed the highest correlation with model performance, suggesting that this phenomenon is objective and not specific to any one model.

\subsection{Sensitivity Analysis of Different Structural Features}
\label{subsec3}
\begin{figure*}[!t]
\captionsetup{aboveskip=0pt,  
                belowskip=-18pt}  
  \centering
  \includegraphics[width=\linewidth]{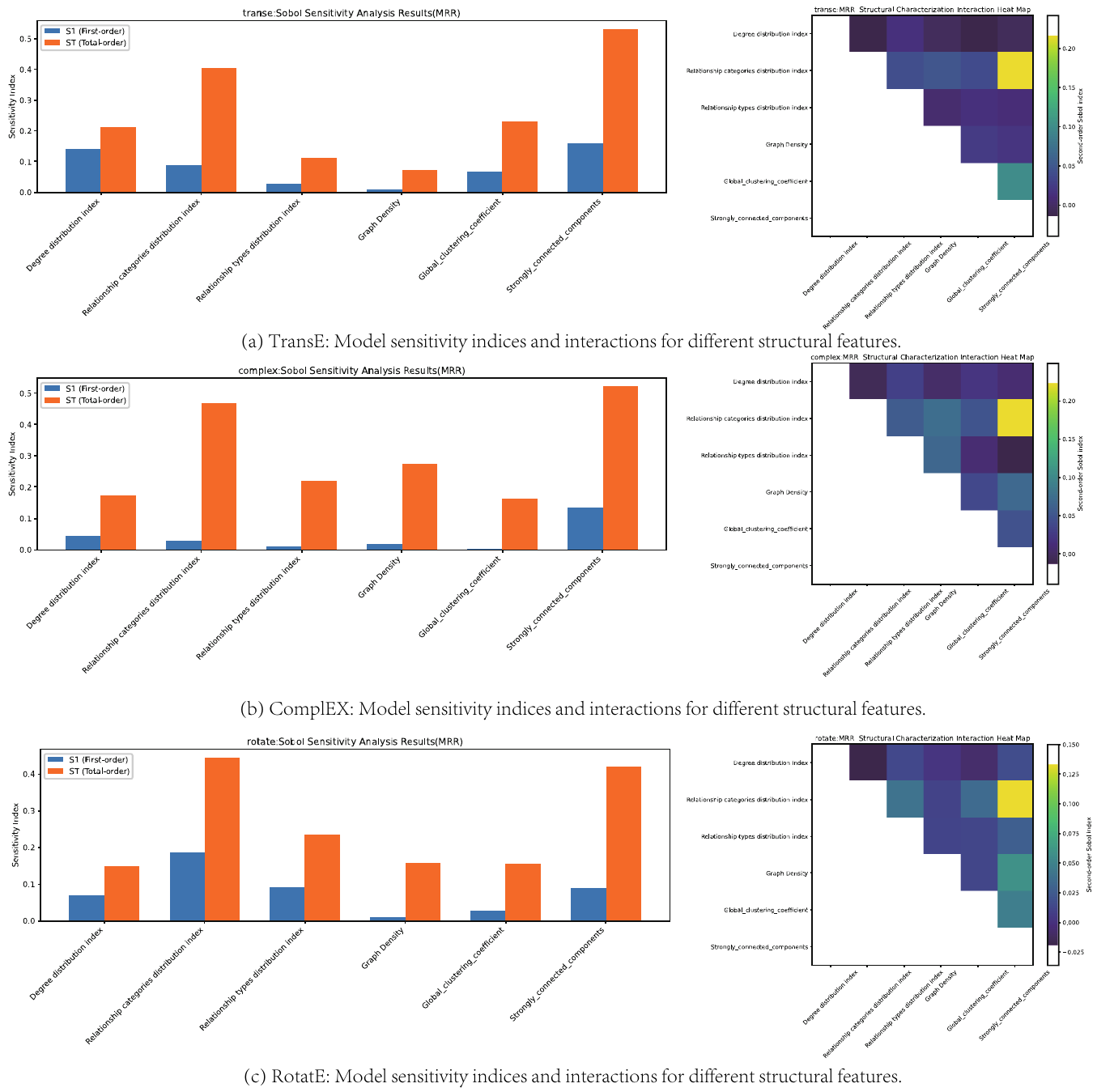}
  \caption{Plot of results of sobol sensitivity analysis for different models.}
  \label{fig3}
\end{figure*}

Through the correlation analysis, we observed a negative correlation between the relationship category distribution index and model performance. To further quantify the impact of different structural features on model performance, this paper uses Sobol sensitivity analysis~\cite{ref38} to conduct a quantitative study of the influence of the aforementioned structural features on model performance. The core idea of Sobol sensitivity analysis is to evaluate the contribution of each input variable (or its interactions) to the output result based on variations in the input variables. This is done by decomposing the output variance to quantify the influence of the input variables. In this study, subgraphs with different structural features are obtained through subgraph sampling, thereby altering the input variables to evaluate the influence of each structural feature on performance. Additionally, considering the interactions between different structural features, the second-order indices from Sobol sensitivity analysis are used to generate a heatmap of the interactions between structural features. The results are shown in Figure~\ref{fig3},  where \( S_1 \) represents the first-order sensitivity index, which indicates the individual impact of an input variable on the output result, measuring the direct contribution of the variable’s variation to the output. A larger \( S_1 \) value suggests a greater contribution of that variable. \( S_T \) represents the total sensitivity index, which measures the total contribution of the input variable and its interactions with other input variables to the output result.

From the result plot, it can be seen that the relationship category distribution index and the number of strongly connected components have a significant impact on different models. Combined with the previous correlation analysis, we can conclude that the distribution of relationship categories not only has a strong correlation with model performance but also has a substantial impact on model performance. From the second-order heatmap, it is observed that the relationship category distribution index not only significantly impacts link prediction performance by itself but also interacts with other structural features to influence the model’s prediction results. The most significant interaction occurs between the relationship category distribution index and the number of strongly connected components, further reinforcing the conclusion that the distribution of relationship categories in the dataset is a key factor affecting model performance. It is noteworthy that previous studies have suggested that the long-tail distribution phenomenon caused by the imbalance in the distribution of relationship types significantly affects model performance. However, our research has shown that on FB15k-237, the impact of relationship type distribution is far less significant than that of the relationship category distribution and the number of strongly connected components.

\subsection{Mechanism Study of Relationship Categories Distribution's Impact on Modeling}
\label{subsec4}
Although many studies have previously recognized that the distribution of relationship categories affects model performance and have made improvements accordingly, to the best of our knowledge, no systematic analysis of its impact has been conducted. In the previous sections, through correlation and sensitivity analyses, we have demonstrated the critical impact of the relationship category distribution on model performance. In the following two sections, we will explore the mechanisms by which it influences model performance from multiple perspectives. This section will analyze the extent to which different relationship categories' data are affected by parameter changes. Both this section and the next will use the TransE model for experiments.

\begin{figure*}[!t]
  \captionsetup{aboveskip=0pt,  
                belowskip=-12pt}  
  \centering
  \includegraphics[width=\linewidth]{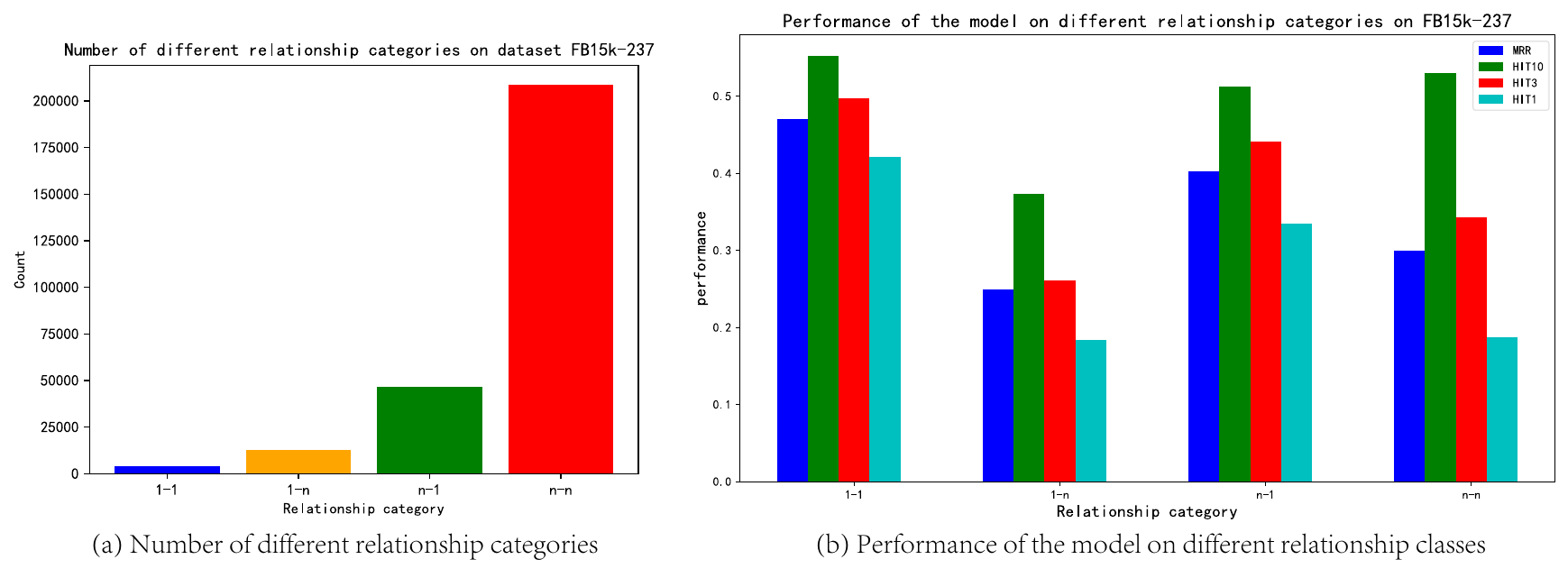}
  \caption{Distribution of number and performance of different relationship categories.}
  \label{fig4}
\end{figure*}

First, we performed a statistical analysis of the distribution of each relationship category in the FB15k-237 dataset. We then tested the model's performance on different relationship categories, as shown in Figure~\ref{fig4}. From the results, it can be observed that the relationship category distribution in the FB15k-237 dataset is highly imbalanced, with the n-n category accounting for more than 70\%, while other relationship categories have fewer instances. More significantly, the model's prediction performance on the n-n category is not ideal. This suggests that subgraphs with lower performance may have a higher proportion of the n-n relationship category. Therefore, a key area for improving model performance in the future could focus on enhancing the model's ability to learn from the n-n category relationships.

\begin{figure*}[!t]
  \captionsetup{aboveskip=2pt,  
                belowskip=-16pt}  
  \centering
  \includegraphics[width=\linewidth]{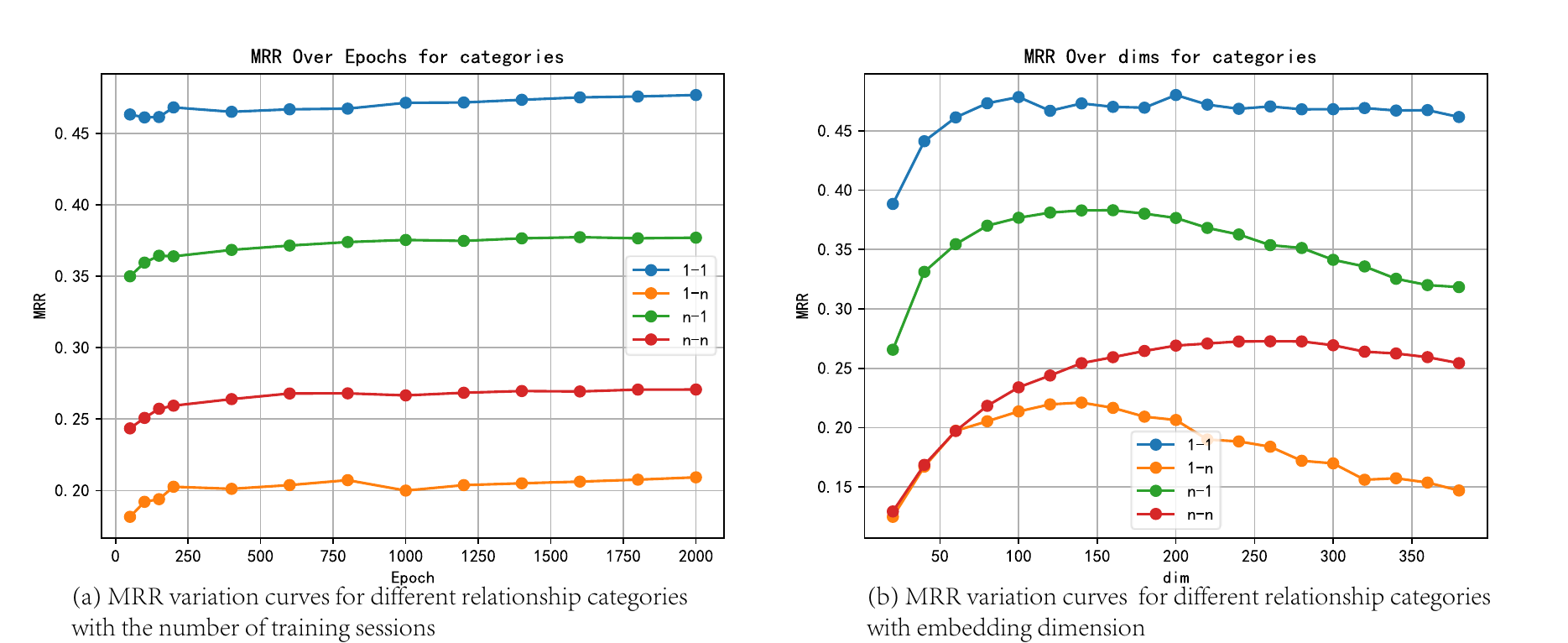}
  \caption{Curve of performance of different relation classes with training time, embedding dimension.}
  \label{fig5}
\end{figure*}

Next, to explore the impact of hyperparameters on the model's performance across different relationship categories, we set different training times and embedding dimensions to test the performance on various relationship categories and plotted the corresponding change curves. The results are shown in Figure~\ref{fig5}.

\begin{enumerate}
    \item \textbf{Model Performance on Different Relationship Categories as a Function of Training Time (Epochs):} For the performance variation of different relationship categories with training time, it can be observed that relationship categories with more training samples, such as the n-n category, require longer training times to reach a state of convergence. However, this does not lead to a significant improvement in model performance on relationship categories with more training samples.
    \item \textbf{Model Performance on Different Relationship Categories as a Function of Embedding Dimensions (Dims):} Comparing the training performance of different relationship categories at various embedding dimensions, relationship categories with larger datasets and more complex structures require higher embedding dimensions to effectively learn and fit the data. This may be due to the complexity of the semantic patterns in these relationship categories, and an increase in embedding dimensions helps the model learn more granular feature representations. For simpler relationship categories, the model can achieve convergence with lower embedding dimensions, indicating lower learning difficulty. However, as embedding dimensions increase, these relationship categories are more prone to overfitting because their training samples are insufficient to support a complex model at higher dimensions. Therefore, in the future, the model can allocate different resources to different relationship categories, enabling it to better learn complex relationship category information while avoiding overfitting in simpler categories.
\end{enumerate}

In summary, we found that the training difficulty and resource allocation differ across relationship categories. The n-n category, which has a large dataset and complex structure, is a key factor hindering the improvement of model performance. Future research can focus on enhancing the model's ability to learn from complex relationship categories and more reasonably allocate training resources to avoid overfitting in simpler relationship categories.

\subsection{Mechanism Explanation of the Impact of Relationship Category Distribution on Model Performance}
\label{subsec5}
\begin{figure*}[!t]
  \captionsetup{aboveskip=0pt,  
                belowskip=-16pt}  
  \centering
  \includegraphics[width=\linewidth]{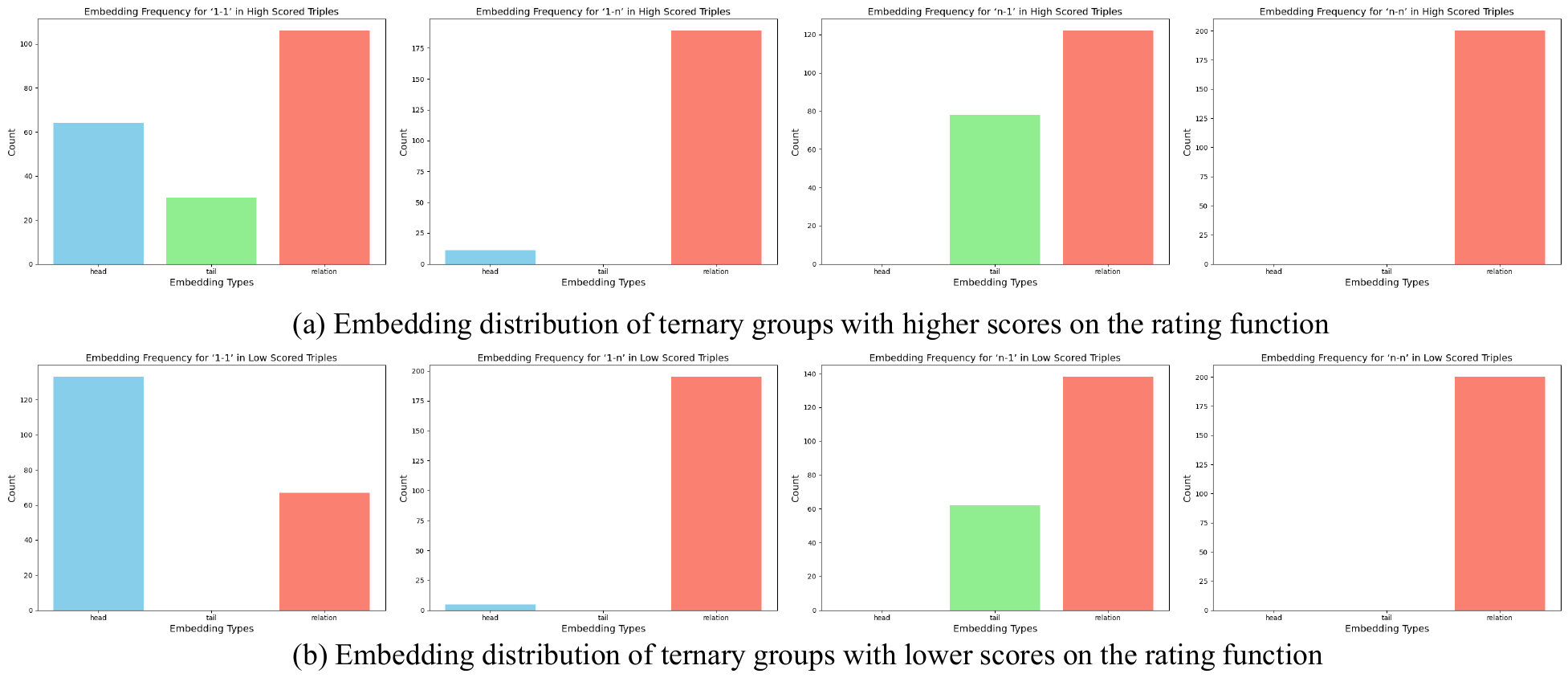}
  \caption{Distribution of embedding vectors for higher and lower scoring ternary groups under the scoring function.}
  \label{fig6}
\end{figure*}

Finally, this paper uses the LIME model to analyze the pathways through which different relationship categories affect model performance. The LIME model is a model-agnostic explanation method designed to interpret the predictions of black-box models. We first divide the dataset by relationship category and apply a scoring function to the triples in each category. Then, we select representative triples with high and low scores and use the LIME model to analyze the sources of these scores. By using the LIME model to compute the importance of entity embedding vectors and relationship embedding vectors to the scoring function, we observe how the relationship categories influence model performance through embedding vectors. The results are shown in Figure~\ref{fig6}.

Based on the analysis results from the LIME model, we find that different relationship categories distinctly influence embedding vectors; nevertheless, a commonality is the consistently significant role of relationship embeddings across all categories. This indicates that relationship embeddings are crucial for capturing relationships among entities within knowledge graphs, emphasizing the need to prioritize their optimization during model design. The specific impacts of relationship categories on embeddings are as follows: for the 1-1 relationship category, triples with high scores exhibit similar importance for both head and tail entity embeddings, suggesting balanced embedding requirements. For 1-n categories, head entity embeddings have greater significance, whereas for n-1 categories, tail entity embeddings dominate. Thus, when handling 1-n or n-1 relationship categories, model designs should emphasize the respective embeddings of head or tail entities. Meanwhile, the n-n relationship category heavily relies on the importance of relationship embeddings, indicating that capturing complex bidirectional interactions is critical for embedding representation in such scenarios.

These findings indicate that different relationship categories indirectly affect link prediction performance by changing the importance of entity embeddings and relationship embeddings within the model. Future work can leverage distinct embedding strategies tailored to specific relationship categories, especially to enhance modeling capabilities for complex relationships, ultimately improving overall link prediction performance.

\section{Conclusion}

This paper systematically investigates the impact of structural features of the FB15k-237 dataset on the performance of knowledge graph embedding (KGE) models. By employing the proposed subgraph sampling method, we construct subgraphs with varying structural features and conduct correlation and sensitivity analyses. Our experimental results reveal that the primary factor constraining model performance on FB15k-237 is not, as previously speculated, the complexity of relationship patterns or semantic structures, but rather the relationship category distribution features in the dataset, followed by the size of strongly connected components. Similar conclusions were drawn from evaluations on three representative KGE models—TransE, ComplEx, and RotatE—highlight the generality of our findings and further indicate that these performance constraints originate from intrinsic dataset properties rather than the specifics of model architectures. Moreover, in contrast to conventional optimization approaches, which rely on parameter tuning or model stacking strategies through iterative trial-and-error, our study introduces an interpretable mechanism analysis using the LIME model. Through this approach, we elucidate the pathways by which relationship categories impact model performance, showing explicitly how these categories modulate the relative importance of entity embeddings and relationship embeddings, thereby influencing prediction outcomes.

Our research not only empirically confirms the significant effect of relationship category distributions on link prediction performance but also provides theoretical insights into the underlying mechanisms. These insights offer essential theoretical support for future work aimed at enhancing the expressive capability of models concerning complex relationship categories, optimizing resource allocation strategies, and refining embedding techniques. Furthermore, the analytical framework proposed in this paper serves as a generalizable foundation for modeling complex dataset structures and designing efficient embedding mechanisms, potentially driving further advancements in knowledge graph link prediction methodologies.

\section{Acknowledgements}
This work was supported by the National Natural Science Foundation of China (Grant No. U1801262) and the Guangdong Provincial Science and Technology Project (Grant No. 2019B010154003).


\begin{thebibliography}{38}
\bibitem{ref1}
Krizhevsky, A., Sutskever, I., \& Hinton, G. E. (2017). ImageNet classification with deep convolutional neural networks. Communications of the ACM, 60(6), 84-90.

\bibitem{ref2}
He, K., Zhang, X., Ren, S., \& Sun, J. (2016). Deep residual learning for image recognition. In Proceedings of the IEEE conference on computer vision and pattern recognition (pp. 770-778).

\bibitem{ref3}
Wang, A., Singh, A., Michael, J., Hill, F., Levy, O., \& Bowman, S. R. GLUE: A Multi-Task Benchmark and Analysis Platform for Natural Language Understanding. In International Conference on Learning Representations.

\bibitem{ref4}
Devlin, J., Chang, M. W., Lee, K., \& Toutanova, K. (2019, June). Bert: Pre-training of deep bidirectional transformers for language understanding. In Proceedings of the 2019 conference of the North American chapter of the association for computational linguistics: human language technologies, volume 1 (long and short papers) (pp. 4171-4186).

\bibitem{ref5}
Radford, A., Narasimhan, K., Salimans, T., \& Sutskever, I. (2018). Improving language understanding by generative pre-training.

\bibitem{ref6}
Himmelstein, D. S., Lizee, A., Hessler, C., Brueggeman, L., Chen, S. L., Hadley, D., ... \& Baranzini, S. E. (2017). Systematic integration of biomedical knowledge prioritizes drugs for repurposing. elife, 6, e26726.

\bibitem{ref7}
Toutanova, K., \& Chen, D. (2015, July). Observed versus latent features for knowledge base and text inference. In Proceedings of the 3rd workshop on continuous vector space models and their compositionality (pp. 57-66).

\bibitem{ref8}
Bordes, A., Usunier, N., Garcia-Duran, A., Weston, J., \& Yakhnenko, O. (2013). Translating embeddings for modeling multi-relational data. Advances in neural information processing systems, 26.

\bibitem{ref9}
Zhang, Y., Yao, Q., Dai, W., \& Chen, L. (2020, April). AutoSF: Searching scoring functions for knowledge graph embedding. In 2020 IEEE 36th International Conference on Data Engineering (ICDE) (pp. 433-444). IEEE.

\bibitem{ref10}
Li, Q., Zhong, Y., \& Qin, Y. (2024, November). MoCoKGC: Momentum Contrast Entity Encoding for Knowledge Graph Completion. In Proceedings of the 2024 Conference on Empirical Methods in Natural Language Processing (pp. 14940-14952).

\bibitem{ref11}
Wang, Q., Mao, Z., Wang, B., \& Guo, L. (2017). Knowledge graph embedding: A survey of approaches and applications. IEEE transactions on knowledge and data engineering, 29(12), 2724-2743.

\bibitem{ref12}
Zhang, Z., Guan, Z., Zhang, F., Zhuang, F., An, Z., Wang, F., \& Xu, Y. (2023, July). Weighted knowledge graph embedding. In Proceedings of the 46th international ACM SIGIR conference on research and development in information retrieval (pp. 867-877).

\bibitem{ref13}
Ji, G., He, S., Xu, L., Liu, K., \& Zhao, J. (2015, July). Knowledge graph embedding via dynamic mapping matrix. In Proceedings of the 53rd annual meeting of the association for computational linguistics and the 7th international joint conference on natural language processing (volume 1: Long papers) (pp. 687-696).

\bibitem{ref14}
Lin, Y., Liu, Z., Sun, M., Liu, Y., \& Zhu, X. (2015, February). Learning entity and relation embeddings for knowledge graph completion. In Proceedings of the AAAI conference on artificial intelligence (Vol. 29, No. 1).

\bibitem{ref15}
Yang, B., Yih, S. W. T., He, X., Gao, J., \& Deng, L. (2015, May). Embedding Entities and Relations for Learning and Inference in Knowledge Bases. In Proceedings of the International Conference on Learning Representations (ICLR) 2015.

\bibitem{ref16}
Trouillon, T., Dance, C. R., Gaussier, É., Welbl, J., Riedel, S., \& Bouchard, G. (2017). Knowledge graph completion via complex tensor factorization. Journal of Machine Learning Research, 18(130), 1-38.

\bibitem{ref17}
Balazevic, I., Allen, C., \& Hospedales, T. (2019, November). TuckER: Tensor Factorization for Knowledge Graph Completion. In 2019 Conference on Empirical Methods in Natural Language Processing and 9th International Joint Conference on Natural Language Processing (pp. 5184-5193). Association for Computational Linguistics.

\bibitem{ref18}
Dettmers, T., Minervini, P., Stenetorp, P., \& Riedel, S. (2018, April). Convolutional 2d knowledge graph embeddings. In Proceedings of the AAAI conference on artificial intelligence (Vol. 32, No. 1).

\bibitem{ref19}
Chen, S., Liu, X., Gao, J., Jiao, J., Zhang, R., \& Ji, Y. (2021, November). HittER: Hierarchical Transformers for Knowledge Graph Embeddings. In Proceedings of the 2021 Conference on Empirical Methods in Natural Language Processing (pp. 10395-10407).

\bibitem{ref20}
Zhu, Z., Zhang, Z., Xhonneux, L. P., \& Tang, J. (2021). Neural bellman-ford networks: A general graph neural network framework for link prediction. Advances in neural information processing systems, 34, 29476-29490.

\bibitem{ref21}
Zhang, Y., Zhou, Z., Yao, Q., \& Li, Y. (2022, May). Efficient hyper-parameter search for knowledge graph embedding. In Proceedings of the 60th Annual Meeting of the Association for Computational Linguistics (Volume 1: Long Papers) (pp. 2715-2735).

\bibitem{ref22}
Zhang, Z., Zhuang, F., Zhu, H., Li, C., Xiong, H., He, Q., \& Xu, Y. (2021). Towards robust knowledge graph embedding via multi-task reinforcement learning. IEEE Transactions on Knowledge and Data Engineering, 35(4), 4321-4334.

\bibitem{ref23}
Wang, P., Agarwal, K., Ham, C., Choudhury, S., \& Reddy, C. K. (2021, April). Self-supervised learning of contextual embeddings for link prediction in heterogeneous networks. In Proceedings of the web conference 2021 (pp. 2946-2957).

\bibitem{ref24}
Yang, H., Zhang, L., Su, F., \& Pang, J. (2022, April). What Affects the Performance of Models? Sensitivity Analysis of Knowledge Graph Embedding. In International Conference on Database Systems for Advanced Applications (pp. 698-713). Cham: Springer International Publishing.

\bibitem{ref25}
Wang, Z., Zhang, J., Feng, J., \& Chen, Z. (2014, June). Knowledge graph embedding by translating on hyperplanes. In Proceedings of the AAAI conference on artificial intelligence (Vol. 28, No. 1).

\bibitem{ref26}
Vu, T., Nguyen, T. D., Nguyen, D. Q., \& Phung, D. (2019, June). A Capsule Network-based Embedding Model for Knowledge Graph Completion and Search Personalization. In Proceedings of the 2019 Conference of the North American Chapter of the Association for Computational Linguistics: Human Language Technologies, Volume 1 (Long and Short Papers) (pp. 2180-2189).

\bibitem{ref27}
Lloyd, O., Liu, Y., \& R. Gaunt, T. (2023). Assessing the effects of hyperparameters on knowledge graph embedding quality. Journal of big Data, 10(1), 59.

\bibitem{ref28}
Akrami, F., Saeef, M. S., Zhang, Q., Hu, W., \& Li, C. (2020, June). Realistic re-evaluation of knowledge graph completion methods: An experimental study. In Proceedings of the 2020 ACM SIGMOD International Conference on Management of Data (pp. 1995-2010).

\bibitem{ref29}
Zhang, S., Zhang, J., Song, X., Adeshina, S., Zheng, D., Faloutsos, C., \& Sun, Y. (2023, April). PaGE-Link: Path-based graph neural network explanation for heterogeneous link prediction. In Proceedings of the ACM Web Conference 2023 (pp. 3784-3793).

\bibitem{ref30}
Ma, T., Tao, W., Li, M., Zhang, J., Pan, X., Lin, J., \& Song, B. (2024). KGExplainer: Towards Exploring Connected Subgraph Explanations for Knowledge Graph Completion. arXiv preprint arXiv:2404.03893.

\bibitem{ref31}
Bai, Y., Zhang, B., Xu, N., Zhou, J., Shi, J., \& Diao, Z. (2023). Vision-based navigation and guidance for agricultural autonomous vehicles and robots: A review. Computers and Electronics in Agriculture, 205, 107584.

\bibitem{ref32}
Debowska, A., Boduszek, D., Ochman, M., Hrapkowicz, T., Gaweda, M., Pondel, A., \& Horeczy, B. (2024). Brain Fog Scale (BFS): scale development and validation. Personality and Individual Differences, 216, 112427.

\bibitem{ref33}
Martin, A. J. F., \& Conway, T. M. (2025). Using the Gini Index to quantify urban green inequality: A systematic review and recommended reporting standards. Landscape and Urban Planning, 254, 105231.

\bibitem{ref34}
Ribeiro, M. T., Singh, S., \& Guestrin, C. (2016, August). " Why should i trust you?" Explaining the predictions of any classifier. In Proceedings of the 22nd ACM SIGKDD international conference on knowledge discovery and data mining (pp. 1135-1144).

\bibitem{ref35}
X. Han, S. Cao, X. Lv, et al., ``Openke: An open toolkit for knowledge embedding,'' in \textit{Proc. 2018 Conf. Empir. Methods Nat. Lang. Process: System Demonstrations}, pp. 139--144, 2018.

\bibitem{ref36}
Li, Z., Yang, Y., Li, L., \& Wang, D. (2023). A weighted Pearson correlation coefficient based multi-fault comprehensive diagnosis for battery circuits. Journal of Energy Storage, 60, 106584.

\bibitem{ref37}
Ali Abd Al-Hameed, K. (2022). Spearman's correlation coefficient in statistical analysis. International Journal of Nonlinear Analysis and Applications, 13(1), 3249-3255.

\bibitem{ref38}
Renardy, M., Joslyn, L. R., Millar, J. A., \& Kirschner, D. E. (2021). To Sobol or not to Sobol? The effects of sampling schemes in systems biology applications. Mathematical biosciences, 337, 108593.





\end{thebibliography}



\end{document}